\documentclass[conference]{IEEEtran}
\usepackage{array}
\usepackage{cite}
\usepackage{url}
\usepackage{hyperref}
\usepackage{graphics}
\usepackage{graphicx}
\usepackage{color}
\usepackage{balance}
\usepackage{tabularx}
\usepackage{subfig}

\begin{document}
\title{Facebook's Advertising Platform: New Attack Vectors
 and the Need for Interventions}

\author{\IEEEauthorblockN{Irfan Faizullabhoy }
\IEEEauthorblockA{University of Southern California\\
Email: faizulla@usc.edu}
\and
\IEEEauthorblockN{Aleksandra Korolova}
\IEEEauthorblockA{University of Southern California\\
Email: korolova@usc.edu}
}

\maketitle

\begin{abstract}
Ad targeting is getting more powerful with introduction of new tools, such as Custom Audiences, behavioral targeting, and Audience Insights. Although this is beneficial for businesses as it enables people to receive more relevant advertising, the power of the tools has downsides. In this paper, we focus on three downsides: privacy violations, microtargeting (i.e., the ability to reach a specific individual or individuals without their explicit knowledge that they are the only ones an ad reaches) and ease of reaching marginalized groups. Using Facebook's ad system as a case study, we demonstrate the feasibility of such downsides. We then discuss Facebook's response to our responsible disclosures of the findings and call for additional policy, science, and engineering work to protect consumers in the rapidly evolving ecosystem of ad targeting.
\end{abstract}
\IEEEpeerreviewmaketitle

\section{Introduction}
In the past several years, advertisement platform providers such as Facebook, Google, Twitter, and Pinterest, have developed and released a suite of new tools for advertisers. Those tools leverage the consumer information gathered by the platforms to deliver advertisements more effectively, i.e., help advertisers in targeting or reaching an audience who may be interested in their ads, as determined by the advertiser and the ad platform's algorithms. Although these tools have been a boon for advertisers, each new tool and feature potentially brings new threats to consumer privacy and the welfare of society. 

\textit{Personally Identifying Information Audiences} (PII Audiences~\cite{venkatadri-2018-targeting}), also referred to as Custom Audiences by Facebook~\cite{fb-custom-audience}, Customer Match Audiences by Google~\cite{google-customer-match-policy}, or Tailored Audiences by Twitter~\cite{twitter-tailored-audience}, are an example of one such tool. PII Audiences are advertisement audiences that are created by uploading individuals' personal information, such as email, full name, age, zip code, in order to then deliver advertisements to their associated social media accounts. This mechanism is intended to allow advertisers to perform remarketing across different platforms and to bridge the gap between offline and online interactions~\cite{fb-custom-audience}. For instance, an advertiser is able to send targeted advertisements to customers who visited their store-front and wrote down their name and email on a sign-in list. Although being able to reach audiences via PII Audience creation may be very useful for advertisers, it is not hard to imagine that such audience reach or targeting capabilities can also be misused for violating privacy. For example,~\cite{venkatadri-2018-targeting} have recently showed how PII Audience technology on Facebook's advertising platform could be used to de-anonymize website visitors, and uncover any user's phone number. 

\begin{figure*}[ht]
  \centering
  \begin{subfloat}
  	{\includegraphics[width=0.45\textwidth]{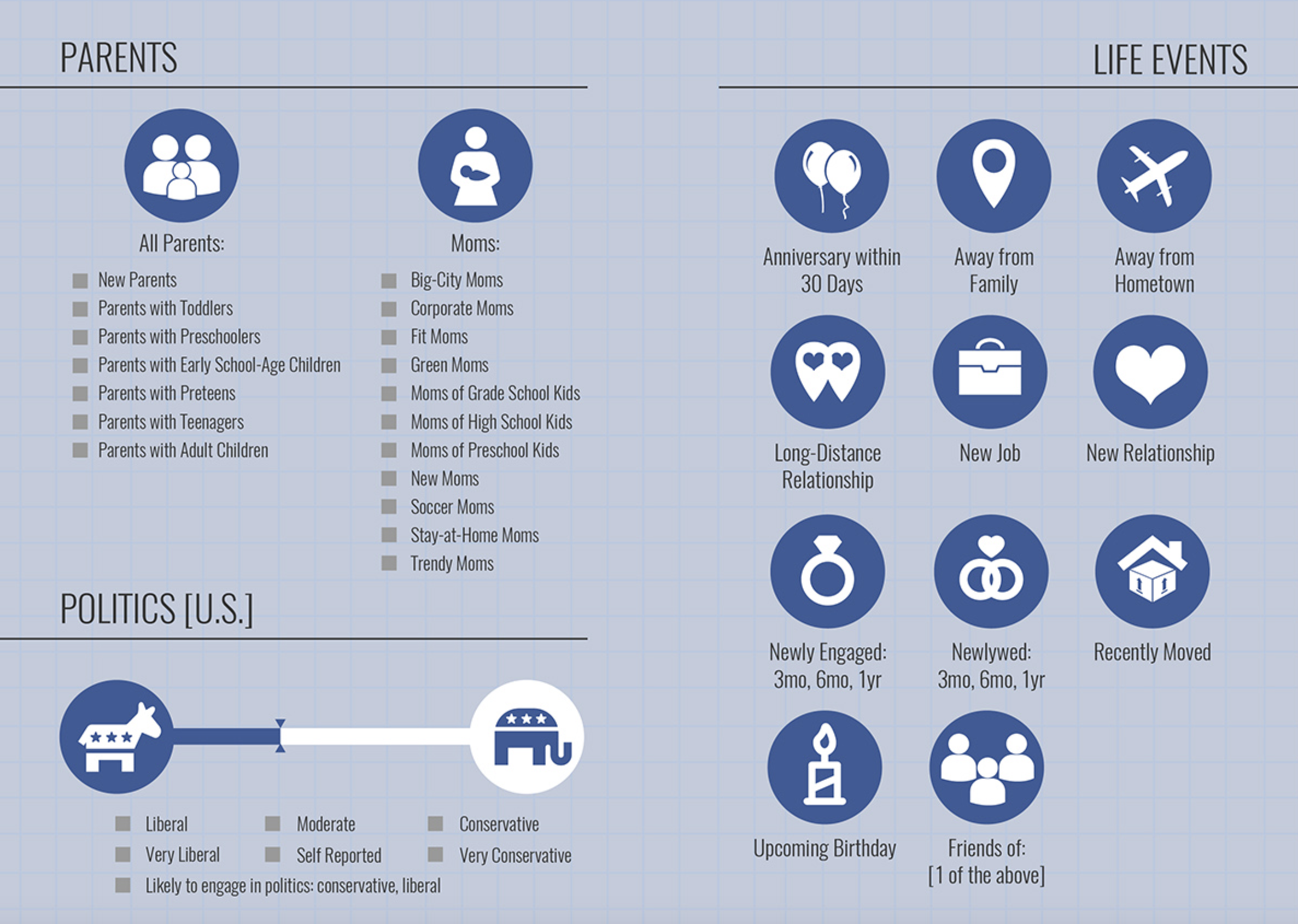}}
  \end{subfloat}
  \qquad
  \begin{subfloat}
  	{\includegraphics[width=0.45\textwidth]{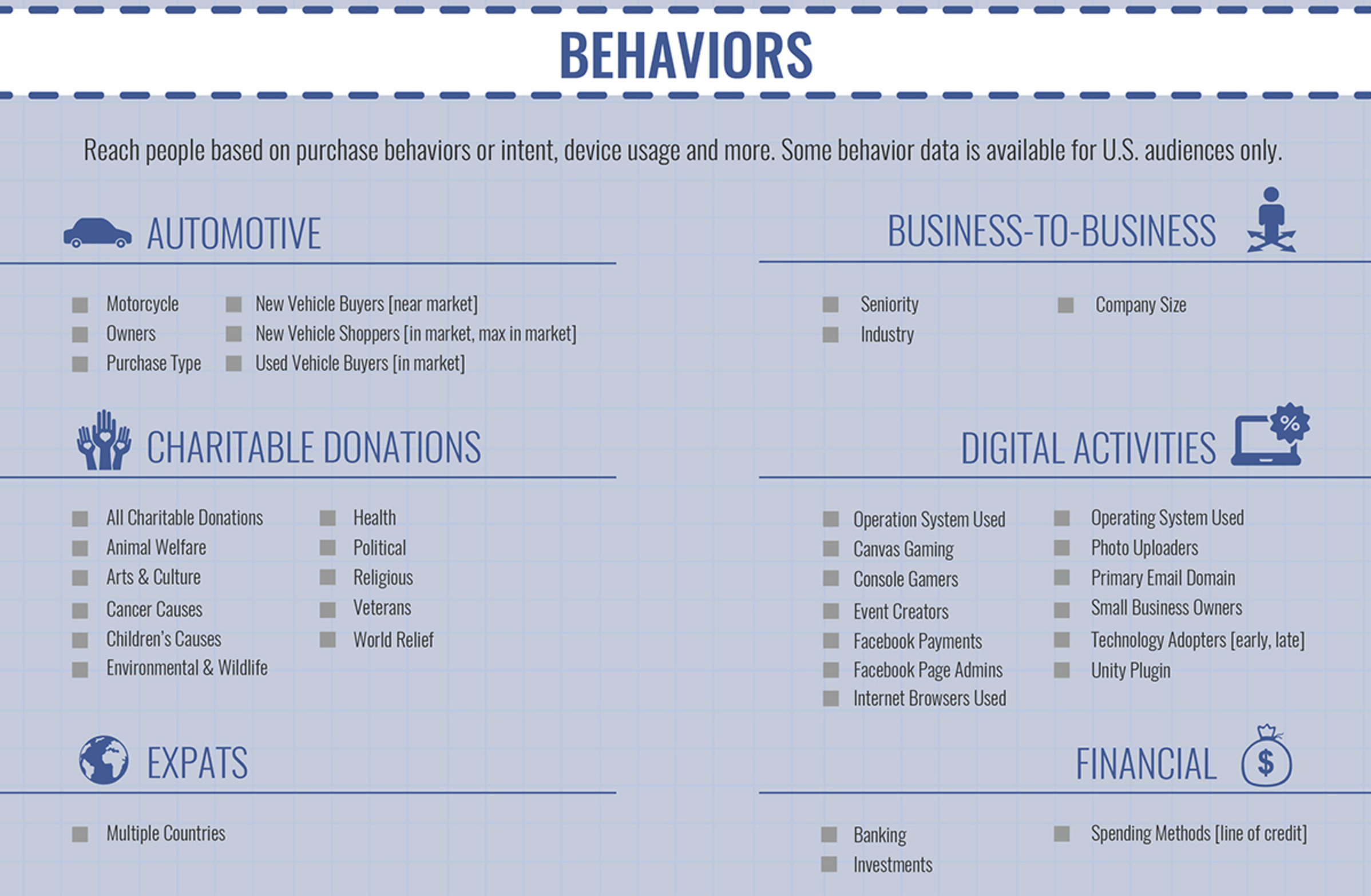}}
  \end{subfloat}
  \caption{Targeting Criteria Infographic \cite{fb-targeting-infographic}}
  \label{fig:infograph}
\end{figure*}

Another emerging advertising technology is \textit{demographic and behavioral targeting}. Social media websites such as Facebook, Google, and Pinterest record and learn from user behavior, taking into account their activity such as location, post content, likes, and self-reported (potentially private) information, to create an internal representation of a user's demographic, interests, and behaviors. The platforms allow advertisers to specify demographic and behavioral targeting criteria which they then match to their learned profile information about the users. For example, advertisers can specify that their ad campaign should target ``Females living in London interested in Shopping between the ages of 18-24," which means that their ad should be shown to users that the ad platform thinks fits these criteria -- a few examples of such targeting criteria are given in Figure \ref{fig:infograph} \cite{fb-targeting-infographic}. 

Again, such detailed targeting capabilities can be abused; for example, by specifying a combination of criteria that match only one individual, the ad campaign can single her out and learn additional information about her~\cite{korolova2011}. The typical protection put in place by ad platforms to prevent this kind of attack is a threshold on the minimum number of users who need to satisfy the targeting criteria before a campaign can be run (Section \ref{sec:others}). In addition to privacy violations, recent work by~\cite{angwin_tobin_varner} has shown that demographic and behavioral targeting may also lead to illegal or discriminatory practices, such as excluding individuals of certain race from seeing housing ads.

\begin{figure*}[ht]
\caption{Audience Insights \cite{fb-audience-insight}}
\centering
\includegraphics[width=0.75\textwidth]{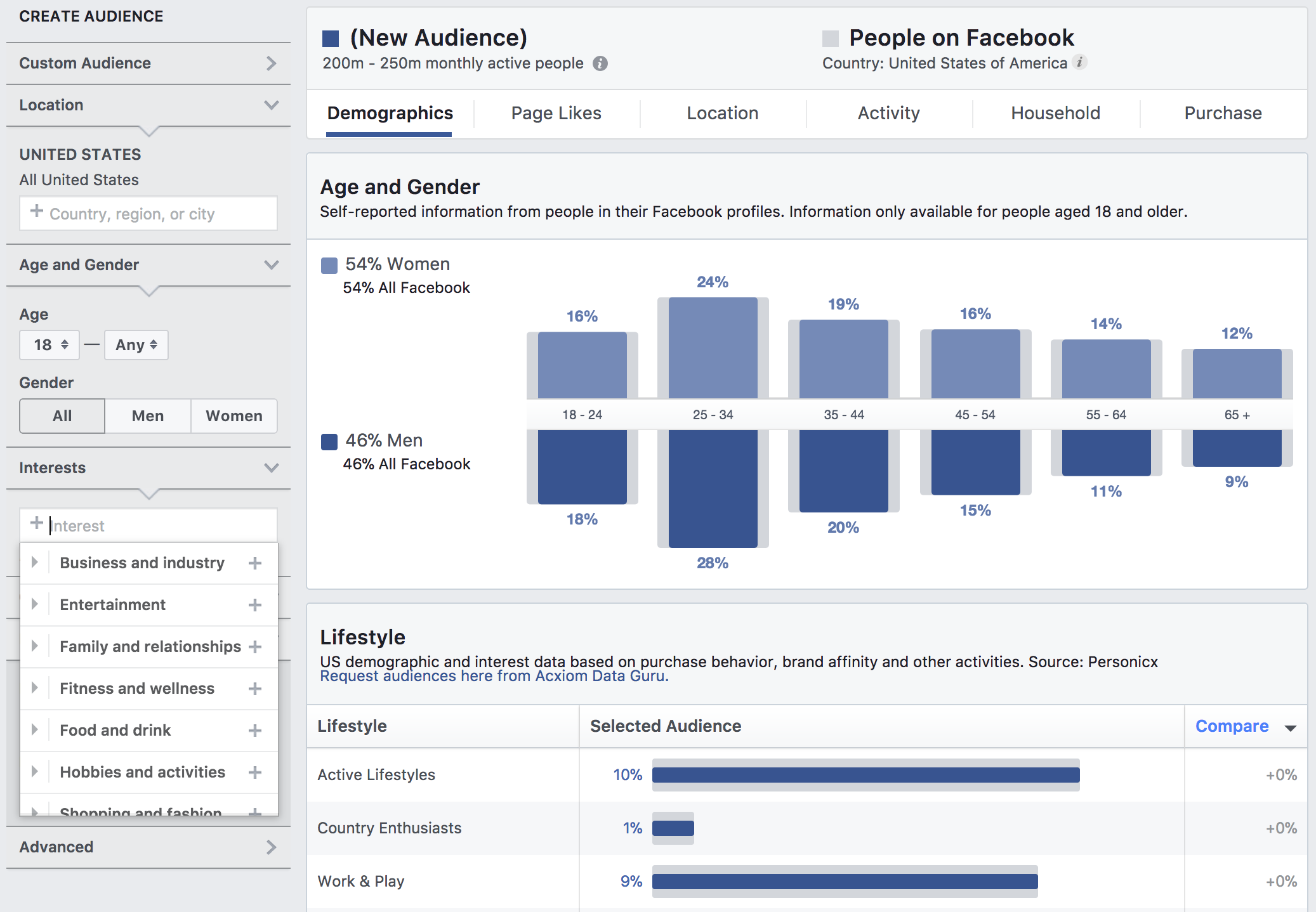}
\label{fig:insights}
\end{figure*}

Finally, another recent tool, is \textit{Audience Insights}, a collection of information reported to advertisers about the users who were reached by their ad. A typical audience insights dashboard goes far beyond the number of people who have seen an ad, and includes information about their gender, wealth and age distribution, interests, locations, etc. See an example Audience Insights page from Facebook in Figure~\ref{fig:insights}. The intent of the tool is to describe the characteristics shared by a large number of users reached by the ad; however, if not properly implemented, it can be exploited to learn private information about individuals.
Again, the privacy protections, if any, typically put in place for such a tool are thresholds chosen in an ad-hoc manner. 

In this work, we demonstrate that despite all the attention devoted to possible harms caused by Facebook's advertising platform and Facebook's stated commitments to privacy, the ad platform can still be exploited to violate privacy, and to cheaply and effectively target individuals or marginalized groups (Section~\ref{sec:facebook}). We then describe Facebook's response to our findings and its implications for the privacy, accountability, and transparency of its advertising platform (Section~\ref{sec:response}). We then briefly discuss the differences in practices of other advertising platforms with respect to ad targeting (Section~\ref{sec:others}). We conclude with a call for change: in accountability, transparency and user control in ad targeting practices, in policy scrutiny to such practices, and in engineering and scientific solutions that could enable powerful targeting while provably preserving privacy (Section~\ref{sec:conclusion}).

\section{Privacy Violations, Microtargeting, and Marginalized Group Targeting on Facebook}\label{sec:facebook}
We identified three novel attack vectors:\\

\textbf{A. Single-Person Insights (enabling privacy violations):} Facebook's Audience Insights product can be used to learn highly private information of an individual Facebook user. In particular, given the knowledge of a person's name, email address and/or phone number, it is possible to learn Facebook's estimate of their age, household composition, interests, income, etc. The reason is that Audience Insights can be run on audiences as small as one person, and when run, insights include 2,000+ categories of information.

\textbf{B. Single-Person Targeting (enabling microtargeting):} Facebook's Custom Audience feature can be exploited to run Facebook-approved campaigns aimed at a single user. The reason is that although Facebook applies a minimum threshold on the Custom Audience size, that threshold is very small and can easily be surpassed by including users who are known to use AdBlock in the Custom Audience specification.

\textbf{C. Single-House Location Targeting (enabling marginalized group targeting):} Facebook's location targeting feature can be used to run Facebook-approved ad campaigns that target arbitrarily small locations (as small as a single house). The reason  is that although the location targeting feature enforces a minimum 1-mile radius, it allows an arbitrary combination of 1-mile radius circles that should be included and excluded from the targeting, enabling one to achieve targeting of a single house.

We describe each of these attacks in detail next. We believe that the primary culprit for making them possible is that Facebook's approach to preventing privacy violations using its ad tools is haphazard. With so many different, rapidly changing advertising tools, many presumably developed by different teams within Facebook, it's difficult to keep track and think of all corner cases and possible interactions between the tools that may have privacy implications, unless one does it using a principled, systematic, transparent, and accountable approach.

\subsection{Single-Person Insights}
This attack vector allows the malicious user to gain information about a single person using the Audience Insights feature of Facebook's advertising platform. Although the feature is nominally intended to present insights about a large group of people, our experiments show that it can be exploited to learn information Facebook possesses about a single person. Insights such as \textit{Income, Net Worth, Interested in Hunting, Buys Plus Size Clothing, Housemates}, etc., can be obtained about a Facebook user merely by specifying their name, phone number and/or email address; for attackers that have a Facebook App, specifying the target's Facebook ID suffices. Not only does the information that can be learned include what Facebook promises to protect from users who aren't Friends, but it also extends beyond the information one formally supplies to Facebook, and may reveal to the advertiser the inferences Facebook's algorithms have made about the user. We describe the approach for executing the attack next.

\begin{enumerate}
\item Create a Custom Audience, using a customer file of only one person's information. Supply Facebook adequate information to uniquely identify the target user. This step can be completed programmatically using the Custom Audiences API, or via the graphical user interface. 
\item Wait 4 hours (max 72 hours) for the newly generated Custom Audience to become available in the Audience Insights page. 
\item Although the newly generated Custom Audience will be considered invalid, it will still be available for use in Audience Insights. Adding this Custom Audience as a filter, navigate to the Page Likes Dashboard of the Audience Insights tool. 
\item On the Page Likes Dashboard, Facebook will show 5-10 of the Pages liked by the user. This enables one to verify whether the targeting was successful, since if Facebook was unable to locate the user, or if the user is inactive, or Facebook was unable to get significant information about them, their Page Likes Dashboard will not appear.

\item Sequentially apply filters from the list (some examples are: ``single," ``male," ``lives with housemates" one at a time~\cite{fb-audience-insight}). If the Page Likes Dashboard does not disappear upon applying a filter, one can assume that Facebook's algorithm determines that the given user meets the criteria selected by the filter. This is the crux of how one can infer private information about the user -- the filters are very detailed, span over 2,000 categories, and range from information such as ``net worth" to ``frequency of travel". The only exception to the information leak we have observed are sexual orientation and life events, as the Page Likes Dashboard does not appear when those filters are placed on small audience sizes regardless of whether or not that audience meets the filter's criteria.
\end{enumerate} 
 
We ran an experiment executing the attack described on our Facebook friends and Facebook friends who we requested to unfriend us for the purpose of this experiment. In all cases, the experiments were done after obtaining the individual's consent. For each individual, we recorded the information we learned about them through the Single-Person Insights attack and compared it with their self-report to us or our prior knowledge about them. We were able to obtain highly accurate information about many potentially sensitive aspects of their lives, such as: Net Income, Relationship Status, Home Value, Age (accurate within 1 year), Interests (Hunting, Dieting, etc.), and Frequency of Travel.
 
Not only does the ability to make such inferences violate Facebook's promises to their users as stated in their privacy policy, as the information that the individual has shared with Facebook with ``Friends Only"/``Only Me" designation can be obtained by anyone, but it also violates reasonable privacy expectations, as the Custom Audience Insights release information that may not have ever been explicitly disclosed by the individual to Facebook, only inferred from their behavior or other sources. Hence, using the Custom Audience feature one has the power of Facebook inference and data collection capabilities, with no associated costs. Questions such as, ``is this person/their wife pregnant?" ``how old are their children?" ``do they like to gamble?" ``are they living at home, or with roommates?" ``do they hunt?" can all be answered, efficiently and at no cost, by anyone. 

\subsection{Single-Person Targeting}
Facebook's ad targeting options, specifically the combination of the Custom Audience \cite{fb-custom-audience} feature with other targeting criteria, can be exploited to run Facebook-approved campaigns aimed at a single user. The reason is that although Facebook applies a minimum threshold on the Custom Audience size for delivering advertisements, that threshold is very small (20 people) and can easily be surpassed by a determined attacker by including fake or complicit users, or users who block the delivery of advertisements in the custom audience. Blocking advertisements is trivial using a Chrome extension such as Facebook Adblock~\cite{fb-ad-blocker}.

The outline of the attack is as follows:
\begin{enumerate}
\item Select a group of 19 Facebook users who use Facebook Adblock \cite{fb-ad-blocker}, are not active on Facebook, or whose accounts you know are fake. Enter their information into a CSV file with information fields that uniquely identify them, and upload this CSV file as a Custom Audience.
\item Add the information of the target person to the CSV file, again ensuring that the information is sufficient for Facebook to uniquely identify that person.
\item Create a custom audience from the CSV file, using the information of the 19 complicit accounts, plus the target.
\item Create and run an ad for the created Custom Audience. 
\end{enumerate}
The goal of Step 1 is to minimize the cost of delivering the advertising message to a single person while meeting Facebook's threshold on the minimum audience size required before a campaign can be run. By selecting users using Facebook Adblock, one ensures that the ad is not delivered to their client device, so, they will most certainly not click it, and thus a Pay-Per-Click campaign will not incur costs other than when it reaches the target individual.

When verifying the feasibility of this attack, we leveraged the same base audience (users of Adblock) many times, in custom audiences that differed by only one person. During our testing, we didn't encounter any problems, which suggests that once one is able to create 19 complicit accounts, one can utilize them in many Custom Audiences targeting different people without being flagged.

\subsection{Single-House Location Targeting}
Facebook offers a location targeting feature for the ad campaigns, which can be exploited to target a small geographic area, as small as a single house, despite its threshold of a minimum 1-mile radius for targeting. 
The reason is that Facebook allows an arbitrary combination of 1-mile radius circles that should be included and excluded from the targeting, enabling one to achieve targeting of a single house.

\begin{figure}[tb]
\caption{Precise Location Targeting using Inclusion / Exclusion of Areas}
\centering
\includegraphics[width=0.4\textwidth]{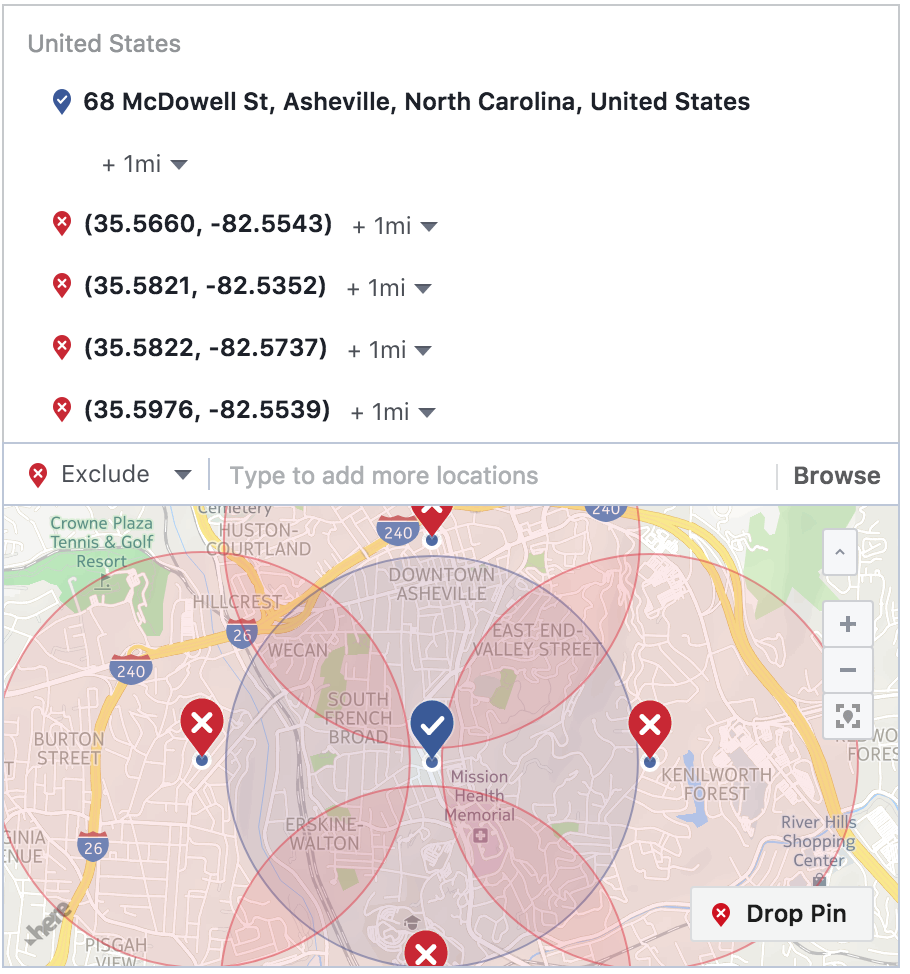}
\label{fig:location}
\end{figure}

The outline of the attack is as follows:
\begin{enumerate}
\item Begin the process for running a Facebook Advertising campaign, and navigate to the ``Target by Location" section.
\item Select the ``Include" option, type in the location of interest (an address), and constrict the radius to 1 mile.
\item Change the option for additional locations to ``Exclude" and drop pins around the perimeter of your original ``Include" Radius, constricting exclusion radii to 1 mile as well. Repeat until you have geo-fenced the area of interest (see Figure~\ref{fig:location} for an example).
\item Run the ad. Facebook will deliver the advertisement as long as there are over 20 users that match your advertising campaign's criteria.
\end{enumerate}
We experimented with the following attack vector on houses in our neighborhood, confirming ad delivery even when the target area was only hundreds of feet wide.

This ad targeting capability makes it easy and cheap to run advertising campaigns targeting specific people or specific vulnerable populations, by simply specifying the geographic location members of that vulnerable population visit, such as a Planned Parenthood clinic, Rehab Center, Cancer Treatment facility, etc. In the pre-Facebook-advertising world, one would have to physically stand outside a Planned Parenthood location in order to deliver a message to the visitors. Now with Facebook, anyone in the world can do so via an ad campaign.

\section{Facebook's Response and Policies}\label{sec:response}
All of the attack vectors mentioned above were reported to Facebook's Responsible Disclosure program upon discovery and confirmation of viability via repeated trials of experiments. Facebook responded to the disclosures with varying levels of concern and promptness. 

\subsection{Single-Person Insights}
We received a response in 4 days that consisted of a request for video proof. Upon providing proof of the attack, Facebook fixed the vulnerability by increasing the threshold necessary for generating the Page Likes Dashboard for Audience Insights on a Custom Audience. Now, only valid Custom Audiences (20 people or more) are shown in the Audience Insights page. Facebook awarded a bug bounty of \$2,000 confirming that ``this may allow a malicious user to infer private information of another user given the person's name, email address and/or phone number."

\subsection{Single-Person Targeting}
Facebook has confirmed that they have received our report. However, for 2 months after the report submission, Facebook did not acknowledge or address it. This is a very slow response time, indicating that preventing Single-Person Targeting is not a priority for Facebook. When Facebook eventually responded, they did not acknowledge a bug or award a bounty, but asked for our suggestions for fixing it, or expectations for how it should work. We suggested that Facebook increase Custom Audience thresholds to levels matching those of other companies, 500 - 1,000 people, making it more difficult for such attacks to be executed. We also recommended that Facebook make additional requirements for which members of the Custom Audience are counted towards the minimum threshold. Facebook has not yet responded to these suggestions.

\subsection{Single-House Targeting}
Facebook's response to the report was to ask to ``clarify how this bug is able to compromise the integrity of Facebook user data, circumvent the privacy protections of Facebook user data, or enable access to a system within Facebook's infrastructure." 

We followed up with a clarification that when extremely precise location targeting, such as targeting of a single house or building, is allowed, it gives the advertiser data about the performance of that campaign. Using the insights and performance data that Facebook delivers, one can see the exact number of people who have seen the ad, what age-group they are in, their gender, and device type~\cite{ad-performance}. Facebook did not respond to this concern and closed the bug bounty report, with no ability for us to reply further. 

\subsection{Discussion of Facebook's Response and Responsible Disclosure Program}
Facebook's response to our Whitehat Reports of ``Single-Person Targeting" that ``This is working as designed" shows an apathy toward microtargeting and circumventions of the rudimentary microtargeting protections Facebook has put in place. Facebook's response to our ``Single-House Targeting" report shows a disregard for the need to limit the ease of targeting marginalized groups. Furthermore, Facebook's advertising and data use policy do not prohibit or discourage microtargeting \cite{fb-ad-policy, facebook-data-use}. The policies do mention that an advertisement cannot display ``implied knowledge" (e.g., first name, ethnicity, financial status) about the target \cite{fb-ad-policy}. However, this clarification does more to help advertisers conceal microtargeted advertisements than to protect individuals.

Additionally, Facebook's Whitehat program policies~\cite{white-hat} make it difficult, and, in some cases, impossible for researchers to discover and report important attack vectors without violating the policies, particularly as it relates to attacks using its advertising platform. In particular, their ``policies only allow testing against test subjects but not normal Facebook users." Not only is this inconvenient, but many of the attack vectors, such as the ``Single-Person Insights" attack could not have been discovered with test users. This is because test users are blank accounts and do not have any private data associated with them, therefore our experiments would not have shown any private information.

\section{Other Big Advertiser Policies}\label{sec:others}
We performed a brief survey of the advertising policies and practices of three other tech companies with advertising platforms. In Table~\ref{table:policies} we present our findings on the minimum thresholds used by them for PII Audiences and their position on microtargeting, based on information we gathered from testing their platforms and reading their relevant advertising and data-use policies \cite{google-customer-match-policy, fb-ad-policy, linkedin-guidelines}.

\vspace{-0.2cm}
\begin{table}[!h]
\caption{}
\label{table:policies}
\begin{center}
\begin{tabular}{| p{1.1cm} | p{1.76cm} | p{4cm}  |  p{0.1cm} |}
\hline
\textbf{Company} & \textbf{PII Audience Threshold} & \textbf{Policy Prohibits Microtargeting?} \\
\hline

Facebook & 20 & No  \\ \hline
Google & 1,000 & Yes  \\ \hline
LinkedIn & 300  & No  \\ \hline
Twitter & 500 & No  \\ \hline

\end{tabular}
\end{center}
\end{table}
\vspace{-0.2cm}

With the exception of Google, no one takes a hard stance on microtargeting in their policies. However, the minimum thresholds enforced by Google, LinkedIn, and Twitter are an order of magnitude larger than Facebook's, showing a more significant effort to prevent it, even if it is not prohibited. In addition, when testing whether or not other companies rigorously enforce their PII thresholds, we found that Twitter does not allow inclusion of spam account (as decided by Twitter), in its Tailored Audiences. This signals that there are varying levels of attention being paid to enforcing minimum thresholds and preventing their circumvention among companies, with Facebook being the least concerned one.

\section{Conclusion}\label{sec:conclusion}
The advent of powerful online ad targeting which is currently non-transparent threatens the well-being of both individuals and society at large.

At the individual level, microtargeting makes stalking and harassment easy and cheap. With only a few cents, an attacker can deliver targeted ads to a particular victim or to a group of people satisfying certain characteristics in minutes. Furthermore, as shown with our ``Single-Person Insights" attack vector and in previous work by~\cite{korolova2011}, an attacker can use microtargeting to gain highly private information on a given user at no cost, and without being Facebook Friends.

At a societal level, ad targeting, particularly one that allows microtargeting and a selection of audience according to arbitrary characteristics, can be used to effectively manipulate public opinion. The most prominent example of this is the digital campaign of Cambridge Analytica, which leveraged highly-targeted and personality-based targeting criteria and content to ``divide and conquer" the public opinion 
via personalized microtargeted advertisements~\cite{nytimes-camb, krogerus_2017}.  
Although political messaging is the most scrutinized area of ad targeting, it is not inconceivable that there are other areas in which powerful ad targeting in the hands of a manipulative entity can have negative consequences, such as health, tolerance to opposing view points, economic habits, education, etc.

Additionally, current and rapidly developing new advertising tools may result in ad campaigns that are discriminatory according to age, sexuality, gender, wealth and even weight with no legal recourse~\cite{angwinage, angwin_tobin_varner, datta2018}. Recent work has also shown that even proper moderation of feature-based targeting (e.g., not allowing targeting by gender for job ads) is insufficient to prevent discrimination, as using new tools, such as Lookalike Audiences \cite{fb-lookalike-audience} and PII Audiences, one can discriminate without explicitly setting discriminatory features~\cite{speicher2018}.

Given the magnitude and urgency of the problems posed, we advocate for the following changes within the context of Facebook's advertising systems:

\subsection{Full Transparency and Empowerment of Consumers}
\subsubsection{Transparency}
We advocate for full transparency of targeted ads: interested users should be able to see who created each advertisement, all of its targeting criteria, and the approximate number of people seeing the ad. 
Currently, Facebook only displays the two most innocuous and public targeting features, usually location and age when there may be more insidious, personal targeting features at play \cite{fb-ad-transparency}.

\subsubsection{Opt Out}
We advocate that users should be able to effortlessly opt out of targeted advertisements, of being included in custom audiences, lookalike audiences, etc.

Currently, an opt out of interest-based advertising based on one's activity requires painstakingly removing each interest individually from the ad preference panel. For active Facebook users, such a panel could include hundreds of interests and thus require non-trivial time effort. Moreover, Facebook does not let users opt out from all targeting categories, e.g., inferred information, such as income range and home value, is not presented to users in their ad preference settings~\cite{fb-ad-transparency}.

\subsubsection{Crowd-sourcing and Accountability}
We advocate for Facebook to give users the tools to meaningfully report suspicious ads and advertisers, and for Facebook to analyze those reports and take appropriate action with the advertiser and report it to the user in near real-time.

\subsection{New Engineering and Scientific Approaches}
Machine learning, which has been successful in areas ranging from NLP to vision, can also play a role in identifying advertisement campaigns and advertisers that aim to microtarget, harass, or discriminate. Software engineering and testing techniques could be brought to ensure that minimum thresholds are applied systematically across all ad tools. Provable privacy techniques such as differential privacy may be useful for making existing ad tools such as Lookalike Audience provably privacy-preserving~\cite{kenthapadi2012privacy} and informing the development of new ones. We believe that close collaboration between ad platform designers and academics with a true dedication to finding new engineering and scientific approaches to the privacy and microtargeting problems could lead to outcomes that would benefit everyone in the ecosystem -- the users, the ad platform designers, the advertisers.

\subsection{Policy Scrutiny}
Finally, we advocate that policy makers and legal scholars get engaged in influencing the practices of ad platforms, particularly in the cases when they can lead to privacy violations and discrimination, via raising awareness of the issues and developing approaches to holding the platforms liable~\cite{datta2018}.

\vspace{-0.01cm}
\balance
\bibliography{conprobib}
\bibliographystyle{IEEEtran}

\end{document}